\documentclass[nofootinbib,prd]{revtex4}%
\usepackage{amsmath}
\usepackage{amsfonts}
\usepackage{amssymb}
\usepackage{graphicx}%
\begin{document}
\title{Cosmological Constant from a Deformation of the Wheeler-DeWitt Equation}
\author{Remo Garattini}
\email{Remo.Garattini@unibg.it}
\affiliation{Universit\`{a} degli Studi di Bergamo,}
\affiliation{Department of Engineering and Applied Sciences}
\affiliation{Viale Marconi 5, 24044 Dalmine (Bergamo) Italy}
\affiliation{I.N.F.N. - sezione di Milano, Milan, Italy}
\author{Mir Faizal}
\email{f2mir@uwaterloo.ca }
\affiliation{Department of Physics and Astronomy,\\ University of Waterloo,  
 \\Waterloo, Ontario, N2L 3G1, Canada,  }
\affiliation{Department of Physics and Astronomy, University of Lethbridge,  
 \\ Lethbridge,  Alberta, T1K 3M4, Canada}
\begin{abstract}
In this paper, we consider the Wheeler-DeWitt equation modified by a
deformation of the second quantized canonical commutation relations. Such
modified commutation relations are induced by a Generalized Uncertainty
Principle. Since the Wheeler-DeWitt equation can be related to a
Sturm-Liouville problem where the associated eigenvalue can be interpreted as
the cosmological constant, it is possible to explicitly relate such an
eigenvalue to the deformation parameter of the corresponding Wheeler-DeWitt
equation. The analysis is performed in a Mini-Superspace approach where the
scale factor appears as the only degree of freedom. The deformation of the
Wheeler-DeWitt equation gives rise to a Cosmological Constant even in absence
of matter fields. As a Cosmological Constant cannot exists in absence of the
matter fields in the undeformed Mini-Superspace approach, so the existence of
a non-vanishing Cosmological Constant is a direct consequence of the
deformation by the Generalized Uncertainty Principle. In fact, we are able to
demonstrate that a non-vanishing Cosmological Constant exists even in the
deformed flat space. We also discuss the consequences of this deformation on
the big bang singularity.

\end{abstract}
\maketitle


\section{Introduction}

It is expected that the geometry of space-time cannot be measured below a
minimum length scale, which is usually taken to be the Planck scale
\cite{l}-\cite{l2}. At this scale it is likely that quantum fluctuations of
the space-time itself come into play, breaking therefore its description as a
smooth manifold \cite{l4}-\cite{l5}. For instance, in string theory, the
minimum length scale is the string length itself. This means that, in
perturbative string theory, it is not possible to probe the space-time below
the string length scale \cite{z4}-\cite{z5}. The appearance of a minimum
measurable length scale has also been studied in the context of loop quantum
gravity \cite{loop}-\cite{loop1}, in noncommutative field theories
\cite{nc}-\cite{cn} and also in black hole physics\cite{bh}-\cite{bh1}. Even
though the existence of such a minimum length scale is predicted from various
different approaches, it is not consistent with the usual Heisenberg
uncertainty principle, which states that the position of a particle can be
measured with arbitrary precision, if its momentum is not measured. This means
that there is no minimum measurable length scale compatible with the usual
Heisenberg uncertainty principle. To accommodate this mismatch, we need to
introduce a Generalized Uncertainty Principle (GUP)\cite{gu}-\cite{ug}. As the
uncertainty principle is closely related to the Heisenberg algebra, the
generalization of the usual Heisenberg uncertainty principle to GUP deforms
the Heisenberg algebra \cite{1h}-\cite{h1}. This in turns modifies the
coordinate representation of the momentum operator, and this new
representation for the momentum operator produces correction terms for all
quantum mechanical phenomena \cite{d1}-\cite{1d}. A more general deformation
of GUP which incorporates the effect of double special relativity
\cite{ds}-\cite{sd}, has also been studied \cite{li}-\cite{il}. This deformed
Heisenberg algebra also has terms proportional to linear powers of momentum.
Motivated from GUP, the full four momentum of a field theory has also been
modified, and the gauge theory corresponding to this deformation of field
theory have been constructed \cite{gt}-\cite{7}. However, it is also possible
to deform the second quantized commutator between the fields in a similar way.
This has been done for the Wheeler-DeWitt (WDW) equation \cite{w}-\cite{1w}.
This deformed WDW equation has also been used for analyzing quantum black
holes \cite{qb}. The third quantization of this deformed WDW equation has also
been studied \cite{t}. In this analysis, the deformation parameter was
analyzed perturbatively.

Motivated by the deformation of Heisenberg algebra by linear terms in momentum
\cite{li}-\cite{il}, a similar deformation of the second quantized commutator
has been studied \cite{w1}. It may be noted that in the deformation of the
first quantized theories, the GUP parameter can be related to the existence of
an intrinsic measurable length scale in space. Such a relation between a
physical phenomena and GUP deformation has not been studied for the second
quantized theories. A remarkable feature of the WDW equation deformed by GUP
is to avoid singularities in space-time \cite{w1}. This is principally due to
the introduction of a minimum limit to the field resolution. Therefore, it is
quite obvious to try to extend this interesting feature to other contexts, for
example, the cosmological constant. Indeed, it may be noted that the WDW
equation is equivalent to a Sturm-Liouville problem and the related eigenvalue
can be interpreted as a cosmological constant \cite{Remo}. In this context the
cosmological constant is a measure of the degeneracy of the only energy
eigenvalue of the WDW equation without matter fields which obeys the following
equation, $\mathcal{H}\Psi=E\Psi,$ with $E=0$. It is true that an exact
solution has been found by Vilenkin in ordinary GR with a factor ordering
equal to $q=-1$\cite{Vilenkin}. However, except this special case, no other
exact solutions has been found in this context. It is for this reason that one
promising procedure is represented by a variational approach, where the WDW
equation can be cast as a vacuum expectation value (VEV). However, in ordinary
GR, no cosmological constant without matter fields can be produced with the
help of a VEV calculation in a Mini-superspace approach\footnote{Note that the
procedure of building a VEV associated to a Sturm-Liouville problem can be
generalized to include electric and magnetic charges\cite{e,m} and also naked
singularities\cite{Naked}.}. It is for this reason, that the same procedure
has been extended to theories outside GR. Some examples are
Ho\v{r}ava-Lifshitz gravity theory \cite{RemoHL}\footnote{For a traditional
approach to the WDW equation, see also Ref.\cite{OBCZ}.}, Varying Speed of
Light (VSL) cosmology \cite{RemoVSL} and Gravity's Rainbow. However, we have
to say that in all these approaches the kinetic part of the WDW equation does
get any higher order functional derivative correction. Since the deformation
of the second quantized commutator produces higher order functional derivative
contributions for the representation of the canonically conjugate variable to
the field variable \cite{w1}, it will also produce higher derivative
corrections for the kinetic part of the WDW equation. Therefore, our purpose
will be the analysis of the cosmological constant problem using the GUP
deformed WDW equation \cite{cosm}-\cite{cosm1}.

The paper is organized as follows: in Section~\ref{p1} we review the basic
elements of the deformed WDW equation, in Section~\ref{p2} we extract the
corresponding WDW equation in the case of FLRW space-time and we setup the
corresponding Sturm-Liouville. In Section~\ref{p0} we analyze the generalized
semiclassical case corresponding to the deformed WDW equation. In
Section~\ref{q1} we study the flat space case as an example of the generalized
semiclassical case. It may be noted that even though the curvature term is
absent, we still get interesting results due to the deformation of the WDW
equation. In Section~\ref{q2} we consider the case when the operator ordering
parameters do not vanish. Finally, we summarize our results in Section
\ref{p5}. Throughout this manuscript we use units in which $\hbar=c=k=1$.

\section{Deformed Wheeler-DeWitt Equation}

In this section, we will study the deformation of the WDW equation.
\label{p1}The standard WDW equation for a Mini-superspace approach for a
homogeneous, isotropic and closed universe is obtained with the help of the
Friedmann-Lema\^{\i}tre-Robertson-Walker (FLRW) metric%
\begin{equation}
ds^{2}=-N^{2}dt^{2}+a^{2}\left(  t\right)  d\Omega_{3}^{2}, \label{FRW}%
\end{equation}
where%
\begin{equation}
d\Omega_{3}^{2}=\gamma_{ij}dx^{i}dx^{j} \label{domega}%
\end{equation}
is the line element on the three-sphere, $N$ is the lapse function and $a(t)$
denotes the scale factor. In this background, the Ricci curvature tensor and
the scalar curvature read simply%
\begin{equation}
R_{ij}=\frac{2}{a^{2}\left(  t\right)  }\gamma_{ij}\qquad\mathrm{and}\qquad
R=\frac{6}{a^{2}\left(  t\right)  }~,
\end{equation}
respectively. The Einstein-Hilbert action in $(3+1)$-dim is%
\begin{equation}
S=\frac{1}{16\pi G}\int_{\Sigma\times I}\mathcal{L}~dt~d^{3}x=\frac{1}{16\pi
G}\int_{\Sigma\times I}N\sqrt{g}\left[  K^{ij}K_{ij}-K^{2}+R-2\Lambda\right]
~dt~d^{3}x~, \label{action}%
\end{equation}
with $\Lambda$ the cosmological constant, $K_{ij}$ the extrinsic curvature and
$K$ its trace. Using the line element, Eq.~$\left(  \ref{FRW}\right)  $, the
above written action, Eq.~$\left(  \ref{action}\right)  $, becomes%
\begin{equation}
S=-\frac{3\pi}{4G}\int_{I}\left[  \dot{a}^{2}a-a+\frac{\Lambda}{3}%
a^{3}\right]  dt~,
\end{equation}
where we have computed the volume associated to the three-sphere, namely
$V_{3}=2\pi^{2}$, and set $N=1$. The canonical momentum reads%
\begin{equation}
\pi_{a}=\frac{\delta S}{\delta\dot{a}}=-\frac{3\pi}{2G}\dot{a}a~,
\end{equation}
and the resulting Hamiltonian density is%
\begin{align}
\mathcal{H}  &  =\pi_{a}\dot{a}-\mathcal{L}\nonumber\\
&  =-\frac{G}{3\pi a}\pi_{a}^{2}-\frac{3\pi}{4G}a+\frac{3\pi}{4G}\frac
{\Lambda}{3}a^{3}~. \label{H0}%
\end{align}
Following the canonical quantization prescription, we promote $\pi_{a}$ to a
momentum operator, setting%
\begin{equation}
\pi_{a}^{2}\rightarrow-a^{-q}\left[  \frac{\partial}{\partial a}a^{q}%
\frac{\partial}{\partial a}\right]  , \label{ordering}%
\end{equation}
where we have introduced a factor order ambiguity $q$. The generalization to
$k=0,-1$ is straightforward. The WDW equation for such a metric is%
\begin{gather}
H\Psi\left(  a\right)  =\left[  -a^{-q}\left(  \frac{\partial}{\partial
a}a^{q}\frac{\partial}{\partial a}\right)  +\frac{9\pi^{2}}{4G^{2}}\left(
a^{2}-\frac{\Lambda}{3}a^{4}\right)  \right]  \Psi\left(  a\right)
\,,\nonumber\\
\left[  -\frac{\partial^{2}}{\partial a^{2}}-\frac{q}{a}\frac{\partial
}{\partial a}+\frac{9\pi^{2}}{4G^{2}}\left(  a^{2}-\frac{\Lambda}{3}%
a^{4}\right)  \right]  \Psi\left(  a\right)  =0. \label{WDW_0}%
\end{gather}
It represents the quantum version of the invariance with respect to time
reparametrization. Nevertheless, when we include higher derivative correction
to the kinetic part of the WDW equation, we need to modify the second
quantized commutator between the field variables and their conjugate momentum
\cite{w}-\cite{1w}. As the GUP can be generalized to include linear
contributions in the momentum \cite{li}-\cite{il}, a similar modification to
the second quantized momentum has been studied \cite{w1}, and this deformation
of the second quantized commutators modified the representation of the
momentum conjugate to field variables. Now if the original undeformed momentum
conjugate to the scalar factor $a$ is $\tilde{\pi}_{a}$, where
\begin{equation}
\tilde{\pi}_{a}=-i\frac{d}{da},
\end{equation}
then the deformed momentum can be written as \cite{w1}
\begin{equation}
\pi_{a}=\tilde{\pi}_{a}(1-\alpha||\tilde{\pi}_{a}||+2\alpha^{2}||\tilde{\pi
}_{a}||^{2})
\end{equation}
and the generalization of the WDW equation for a FLRW metric is deformed to
\begin{equation}
\left[  \tilde{\pi}_{a}^{2}-2\alpha\pi_{a}^{3}+5\alpha^{2}\pi_{a}^{4}+\left(
\frac{3\pi}{2l_{P}^{2}}\right)  ^{2}a^{2}\left(  1-\frac{\Lambda}{3}%
a^{2}\right)  \right]  \Psi\left(  a\right)  =0.
\end{equation}
This deformation of the WDW equation prevents the existence of singularities
\cite{w1}. This is because this deformation modifies the uncertainty principle
as%
\begin{equation}
\Delta a\Delta\pi_{a}=1-2\alpha<\pi_{a}>+4\alpha^{2}<\pi_{a}^{2}>.
\end{equation}
Thus, we obtain a minimum value for the scale factor of the universe, $\Delta
a\geq\Delta a_{min}$. This minimum value for the scale factor of the universe
can prevent the existence of big bang singularity. As we will relate the
deformation of the WDW equation to the existence of the cosmological constant,
this means it might be possible to avoid the big bang singularity because of
the existence of the cosmological constant in our universe.

The introduction of a factor ordering leads to the following form of the
derivative terms%
\begin{align}
\tilde{\pi}_{a}^{2}  &  =-a^{-s}\frac{d}{da}\left(  a^{s}\frac{d}{da}\right)
,\nonumber\\
\tilde{\pi}_{a}^{3}  &  =-ia^{-p}\frac{d}{da}\left(  a^{p-s}\frac{d}%
{da}\left(  a^{s}\frac{d}{da}\right)  \right)  ,\nonumber\\
\tilde{\pi}_{a}^{4}  &  =a^{-u}\frac{d}{da}\left(  a^{u-p}\frac{d}{da}\left(
a^{p-s}\frac{d}{da}\left(  a^{s}\frac{d}{da}\right)  \right)  \right)  .
\end{align}
So, our final WDW equation would become,%
\begin{align}
&  \left[  \tilde{\pi}_{a}^{2}-2\alpha\pi_{a}^{3}+5\alpha^{2}\pi_{a}%
^{4}+\left(  \frac{3\pi}{2l_{P}^{2}}\right)  ^{2}a^{2}\left(  1-\frac{\Lambda
}{3}a^{2}\right)  \right]  \Psi(a)\nonumber\\
&  =-a^{-q}\frac{d}{da}\left(  a^{q}\frac{d}{da}\Psi(a)\right) \nonumber\\
&  +2\alpha ia^{-u}\frac{d}{da}\left(  a^{u-t}\frac{d}{da}\left(  a^{t}%
\frac{d}{da}\right)  \Psi(a)\right)  +5\alpha^{2}a^{-s}\nonumber\\
&  \times\frac{d}{da}\left(  a^{s-r}\frac{d}{da}\left(  a^{r-p}\frac{d}%
{da}\left(  a^{p}\frac{d}{da}\Psi(a)\right)  \right)  \right) \nonumber\\
&  +\left(  \frac{3\pi}{2l_{P}^{2}}\right)  ^{2}a^{2}\left(  1-\frac{\Lambda
}{3}a^{2}\right)  \Psi(a)=0, \label{WDW}%
\end{align}
where we have defined $l_{P}=\sqrt{G}$. Even though this is a very general
deformation of the WDW equation, we will study the only the deformation of the
WDW equation corresponding to quadratic terms. This is because the
modification of first quantized Heisenberg algebra by linear terms is also
very complicated \cite{6}-\cite{7}, and the quadratic deformation are better
understood \cite{gt}-\cite{tg}. So, we will restrict the WDW equation to the
following form%
\begin{equation}
\left[  \mathcal{H}_{1}+\mathcal{H}_{2}+\left(  \frac{3\pi}{2l_{P}^{2}%
}\right)  ^{2}a^{2}\left(  1-\frac{\Lambda}{3}a^{2}\right)  \right]
\Psi(a)=0, \label{WGV}%
\end{equation}
where%
\begin{equation}
\mathcal{H}_{1}=-a^{-q}\frac{d}{da}\left(  a^{q}\frac{d}{da}\right)
\label{H1}%
\end{equation}
and%
\begin{equation}
\mathcal{H}_{2}=5\alpha_{0}^{2}l_{P}^{2}a^{-s}\frac{d}{da}\left(  a^{s-r}%
\frac{d}{da}\left(  a^{r-p}\frac{d}{da}\left(  a^{p}\frac{d}{da}\right)
\right)  \right)  . \label{H2}%
\end{equation}
To further proceed, we have to transform the WDW equation $\left(
\ref{WGV}\right)  $ in a Sturm-Liouville problem. We recall to the reader that
a Sturm-Liouville differential equation is defined by%
\begin{equation}
\frac{d}{dx}\left(  p\left(  x\right)  \frac{dy\left(  x\right)  }{dx}\right)
+q\left(  x\right)  y\left(  x\right)  +\lambda w\left(  x\right)  y\left(
x\right)  =0 \label{SL}%
\end{equation}
and the normalization is defined by%
\begin{equation}
\int_{a}^{b}dxw\left(  x\right)  y^{\ast}\left(  x\right)  y\left(  x\right)
.
\end{equation}
In the case of the FLRW model we have the following correspondence%
\begin{align}
p\left(  x\right)   &  \rightarrow a^{q}\left(  t\right)  \,,\nonumber\\
q\left(  x\right)   &  \rightarrow\left(  \frac{3\pi}{2l_{P}^{2}}\right)
^{2}a^{q+2}\left(  t\right)  \,,\nonumber\\
w\left(  x\right)   &  \rightarrow a^{q+4}\left(  t\right)  \,,\nonumber\\
y\left(  x\right)   &  \rightarrow\Psi\left(  a\right)  \,,\nonumber\\
\lambda &  \rightarrow\frac{\Lambda}{3}\left(  \frac{3\pi}{2l_{P}^{2}}\right)
^{2}\,, \label{corr}%
\end{align}
and the normalization becomes%
\begin{equation}
\int_{0}^{\infty}daa^{q+4}\Psi^{\ast}\left(  a\right)  \Psi\left(  a\right)  .
\label{Norm1}%
\end{equation}
It is a standard procedure, to convert the Sturm-Liouville problem $\left(
\ref{SL}\right)  $ into a variational problem of the form%
\begin{equation}
F\left[  y\left(  x\right)  \right]  =\frac{-\int_{a}^{b}dxy^{\ast}\left(
x\right)  \left[  \frac{d}{dx}\left(  p\left(  x\right)  \frac{d}{dx}\right)
+q\left(  x\right)  \right]  y\left(  x\right)  }{\int_{a}^{b}dxw\left(
x\right)  y^{\ast}\left(  x\right)  y\left(  x\right)  }\, \label{Funct}%
\end{equation}
or equivalently%
\begin{equation}
F\left[  y\left(  x\right)  \right]  =\frac{-\left[  y^{\ast}\left(  x\right)
p\left(  x\right)  \frac{d}{dx}y\left(  x\right)  \right]  _{a}^{b}+\int
_{a}^{b}dxp\left(  x\right)  \left(  \frac{d}{dx}y\left(  x\right)  \right)
^{2}-q\left(  x\right)  y\left(  x\right)  }{\int_{a}^{b}dxw\left(  x\right)
y^{\ast}\left(  x\right)  y\left(  x\right)  }\,,
\end{equation}
with appropriate boundary conditions. If $y\left(  x\right)  $ is an
eigenfunction of $\left(  \ref{SL}\right)  $, then%
\begin{equation}
\lambda=\frac{-\int_{a}^{b}dxy^{\ast}\left(  x\right)  \left[  \frac{d}%
{dx}\left(  p\left(  x\right)  \frac{d}{dx}\right)  +q\left(  x\right)
\right]  y\left(  x\right)  }{\int_{a}^{b}dxw\left(  x\right)  y^{\ast}\left(
x\right)  y\left(  x\right)  }\,,
\end{equation}
is the eigenvalue, otherwise%
\begin{equation}
\lambda_{1}=\min_{y\left(  x\right)  }\frac{-\int_{a}^{b}dxy^{\ast}\left(
x\right)  \left[  \frac{d}{dx}\left(  p\left(  x\right)  \frac{d}{dx}\right)
+q\left(  x\right)  \right]  y\left(  x\right)  }{\int_{a}^{b}dxw\left(
x\right)  y^{\ast}\left(  x\right)  y\left(  x\right)  }\,.
\end{equation}
It is immediate to recognize that the correspondence $\left(  \ref{corr}%
\right)  $ can be applied directly for an ordinary FLRW model without GUP. In
the next section the Sturm-Liouville procedure will be generalized to include
also the GUP correction.

\section{The Cosmological Constant and the GUP deformation}

\label{p2} It may be noted that the value of the cosmological constant depends
on the operator ordering chosen. Thus, our hope is to find a combination of
the operator ordering in such a way to obtain the observed value of the
cosmological constant. Note that in this approach, it also depends on the GUP
parameter which, in the first quantized quantum mechanics, is fixed by the
requirement of space to have an intrinsic minimum measurable length scale.
However, in this paper, we have deformed a second quantized theory. Therefore,
we will relate the GUP parameter to some physical phenomena in the second
quantized theory, which is here represented by the cosmological constant. For
practical purposes, it is better to rescale the cosmological constant
$\tilde{\Lambda}=\Lambda l_{P}^{2}$ and introduce the dimensionless variable
$x=a/l_{P}$, then Eq.$\left(  \ref{WGV}\right)  $ in the Sturm-Liouville form
becomes%
\begin{equation}
\frac{\int dxx^{q+r+s+p}\Psi^{\ast}\left(  x\right)  \left[  \mathcal{H}%
_{1}+\mathcal{H}_{2}+\frac{9\pi^{2}}{4}x^{2}\right]  \Psi\left(  x\right)
}{\int dxx^{q+r+s+p+4}\Psi^{\ast}\left(  x\right)  \Psi\left(  x\right)
}=\frac{3\pi^{2}\tilde{\Lambda}}{4}, \label{VeV}%
\end{equation}
where $\mathcal{H}_{1}$ and $\mathcal{H}_{2}$ are the dimensionless version of
the operators defined in $\left(  \ref{H1}\right)  $ and $\left(
\ref{H2}\right)  $. A crucial point is represented by the choice of the wave
function. The ordinary WDW in GR is represented by Eq. $\left(  \ref{VeV}%
\right)  $ without $\mathcal{H}_{2}$. A proposal for the trial wave function
could be%
\begin{equation}
\Psi\left(  x\right)  =x^{-\frac{q+1}{2}}\exp\left(  -\frac{\beta x^{4}}%
{2}\right)  . \label{Psi0}%
\end{equation}
This form has been tested in Ref.\cite{RemoVSL} and it has not produced any
eigenvalue. The form $\left(  \ref{Psi0}\right)  $ has been considered by
looking at the asymptotic behavior of the original WDW equation without GUP.
Always in Ref.\cite{RemoVSL}, because of the VSL distortion, the form of
$\Psi\left(  x\right)  $ in $\left(  \ref{Psi0}\right)  $ has been modified
into the form%
\begin{equation}
\Psi\left(  a\right)  =a^{-\frac{q+1}{2}}\left(  \beta{a}\right)  ^{-3\alpha
}\exp\left(  -\frac{\beta{a}^{4}}{2}\right)  , \label{Psia}%
\end{equation}
without a rescaling of the scale factor. Note that in $\left(  \ref{Psia}%
\right)  $, it has been introduced a scale factor with a power which is able
to take into account the short distance behavior. When we introduce the GUP
distortion, the effect of $\mathcal{H}_{2}$ introduces a similar behavior but
with a power to the scale factor that we are unable to fix. For this reason,
we will adopt the following trial wave function of the form%
\begin{equation}
\Psi\left(  x\right)  =x^{\frac{\beta-\left(  q+r+p+s+1\right)  }{2}}%
\exp\left(  -\frac{\beta x^{4}}{2}\right)  , \label{Psib}%
\end{equation}
which is suggested by the asymptotic behavior of the WDW equation for a large
scale factor and for the short range behavior we have introduced a power
depending on the variational parameter. With the help of the integrals
calculated in the appendix, Eq.$\left(  \ref{VeV}\right)  $ including GUP
becomes%
\begin{align}
\tilde{\Lambda}\left(  \beta\right)   &  =\frac{4}{3\pi^{2}}\left(
\frac{K_{1}+K_{2}+P_{1}}{P_{2}}\right)  =\frac{4}{3\pi^{2}}\left(
\frac{{{\beta}}^{\frac{1}{2}}\left(  A+4{\beta}\right)  \Gamma\left(
\frac{\beta-2}{4}\right)  }{4{\Gamma\left(  \frac{\beta}{4}\right)  }}\right.
\nonumber\\
&  \left.  \frac{\alpha_{0}^{2}}{\beta-4}\left(  -5B_{1}{{\beta}-}%
40B_{2}{{\beta}^{2}+}240{{\beta}^{3}}\right)  +\frac{9\pi^{2}\Gamma\left(
\frac{2+\beta}{4}\right)  }{{\Gamma\left(  \frac{\beta}{4}\right)  {\beta}%
}^{\frac{1}{2}}}\right)  {.} \label{VeVGUP}%
\end{align}
We demand that%
\begin{equation}
\frac{d\tilde{\Lambda}\left(  \beta\right)  }{d\beta}=0. \label{dLB}%
\end{equation}
It is immediate to recognize that a general analytic solution is difficult to
find. We are therefore led to consider some specific cases. It may be noted
that as the cosmological constant is non-zero in our universe, this means that
we do have a GUP deformation of the Wheeler-DeWitt equation. However, such a
deformation is known to present the existence of singularities \cite{w1}, and
so the existence of the cosmological constant can prevent the existence of the
big bang singularity.

\section{The Generalized Semiclassical Case}

\label{p0}

In ordinary GR, the case in which the factor order parameter $q=0$ is known as
the semiclassical case. If we adopt the same fixing when the GUP\ is present,
Eq.$\left(  \ref{WGV}\right)  $ reduces to%
\begin{equation}
\left[  -\frac{d^{2}}{l_{P}^{2}dx^{2}}+5\alpha_{0}^{2}\frac{d^{4}}{l_{P}%
^{2}dx^{4}}+\left(  \frac{3\pi}{2l_{P}^{2}}\right)  ^{2}l_{P}^{2}x^{2}\left(
1-\frac{\Lambda}{3}l_{P}^{2}x^{2}\right)  \right]  \Psi(x)=0, \label{WDWS}%
\end{equation}
where we have set $q=r=s=p=0$. Therefore, the trial wave function $\left(
\ref{Psib}\right)  $ reduces to%
\begin{equation}
\Psi\left(  x\right)  =x^{\beta}\exp\left(  -\frac{\beta x^{4}}{2}\right)  .
\end{equation}
If we cast Eq.$\left(  \ref{WDWS}\right)  $ into the Sturm-Liouville form, we
obtain%
\begin{equation}
\frac{\int_{0}^{+\infty}dxx^{\beta/2}\exp\left(  -\frac{\beta x^{4}}%
{2}\right)  \left[  -\frac{d^{2}}{dx^{2}}+5\alpha_{0}^{2}\frac{d^{4}}{dx^{4}%
}+\left(  \frac{3\pi}{2}\right)  ^{2}x^{2}\right]  x^{\beta/2}\exp\left(
-\frac{\beta x^{4}}{2}\right)  }{\int_{0}^{+\infty}dxx^{\beta}\exp\left(
-\beta x^{4}\right)  }=\tilde{\Lambda}\frac{3\pi^{2}}{4} \label{SL0}%
\end{equation}
and Eq.$\left(  \ref{VeVGUP}\right)  $ reduces to%
\begin{equation}
\tilde{\Lambda}_{\alpha_{0}}\left(  \beta\right)  =\frac{4}{3\pi^{2}}\left(
\alpha_{0}^{2}{\frac{20{\beta}^{2}\left(  16{\beta}^{2}-40{\beta}-21\right)
}{{\pi}^{2}\left(  1+\beta\right)  \left(  \beta-3\right)  }}+{\frac
{\sqrt{\beta}\left(  16\,{\beta}^{2}+3\left(  3{\pi}^{2}-4\right)  \beta
-9{\pi}^{2}\right)  \Gamma\left(  \beta/4-1/4\right)  }{12{\pi}^{2}%
\Gamma\left(  \beta/4+5/4\right)  }}\right)  . \label{LF}%
\end{equation}
To fix ideas, we can take three values of $\alpha_{0}$: $\alpha_{0}%
=1,\alpha_{0}=100$ and $\alpha_{0}=1000$. By demanding that%
\begin{equation}
\frac{d\tilde{\Lambda}_{\alpha_{0}}\left(  \beta\right)  }{d\beta}=0,
\end{equation}
we find%
\begin{equation}
\left\{
\begin{tabular}
[c]{ccc}\hline
\multicolumn{1}{|c}{$\alpha_{0}$} & \multicolumn{1}{|c}{$\beta_{m}$} &
\multicolumn{1}{|c|}{$\tilde{\Lambda}_{\alpha_{0}}\left(  \beta_{m}\right)  $%
}\\\hline
$1$ & $1.053$ & $32.69$\\
$10$ & $1.017$ & $249.69$\\
$20$ & $1.012$ & $484.86$%
\end{tabular}
\ \ \right.  .
\end{equation}
We can see that the larger is $\alpha_{0}$, the higher is the eigenvalue
$\tilde{\Lambda}_{\alpha_{0}}\left(  \beta_{m}\right)  $.

\section{Flat Space}

\label{q1}

An interesting example of the generalized semiclassical case is the flat space
case. It may be noted that even when curvature is absent, due to the higher
order derivatives, we expect to find non trivial results in the procedure. Now
because of the absence of the curvature term, Eq.$\left(  \ref{SL0}\right)  $
reduces to%
\begin{equation}
\frac{\int_{0}^{+\infty}dxx^{\beta/2}\exp\left(  -\frac{\beta x^{4}}%
{2}\right)  \left[  -\frac{d^{2}}{dx^{2}}+5\alpha_{0}^{2}\frac{d^{4}}{dx^{4}%
}\right]  x^{\beta/2}\exp\left(  -\frac{\beta x^{4}}{2}\right)  }{\int
_{0}^{+\infty}dxx^{\beta}\exp\left(  -\beta x^{4}\right)  }=\tilde{\Lambda
}\frac{3\pi^{2}}{4}%
\end{equation}
and Eq.$\left(  \ref{LF}\right)  $ simplifies to%
\begin{equation}
\tilde{\Lambda}_{\alpha_{0}}\left(  \beta\right)  =\frac{4}{3\pi^{2}}\left(
\alpha_{0}^{2}{\frac{20{\beta}^{2}\left(  16{\beta}^{2}-40{\beta}-21\right)
}{{\pi}^{2}\left(  1+\beta\right)  \left(  \beta-3\right)  }}+{\frac
{\beta^{\frac{3}{2}}\left(  4{\beta-3}\right)  \Gamma\left(  \beta
/4-1/4\right)  }{3{\pi}^{2}\Gamma\left(  \beta/4+5/4\right)  }}\right)  .
\end{equation}
We fix the same values of Eq.$\left(  \ref{SL0}\right)  $ for $\alpha_{0}$ and
by demanding that%
\begin{equation}
\frac{d\tilde{\Lambda}_{\alpha_{0}}\left(  \beta\right)  }{d\beta}=0,
\end{equation}
we find%
\begin{equation}
\left\{
\begin{tabular}
[c]{ccc}\hline
\multicolumn{1}{|c}{$\alpha_{0}$} & \multicolumn{1}{|c}{$\beta_{m}$} &
\multicolumn{1}{|c|}{$\tilde{\Lambda}_{\alpha_{0}}\left(  \beta_{m}\right)  $%
}\\\hline
$1$ & $1.053$ & $29.24$\\
$10$ & $1.017$ & $246.29$\\
$20$ & $1.012$ & $481.46$%
\end{tabular}
\ \ \ \ \right.  .
\end{equation}
We conclude that in a GUP distortion, the presence of the curvature term is
not very relevant since the pattern of the eigenvalues is very close to the
flat case.

\section{Non-Vanishing Parameters}

\label{q2} Now we will analyze the case where $A\neq0,B_{1}\neq0$ and
$B_{2}\neq0$. For this case, we can see what happens for arbitrary choices of
the parameters. We can fix our attention on the following simple setting%
\begin{equation}
q=1,p=1,r=s=0\qquad\Longrightarrow\qquad A=-9,B_{1}=271,B_{2}=32.
\end{equation}
Then Eq.$\left(  \ref{VeVGUP}\right)  $ becomes%
\begin{align}
\tilde{\Lambda}\left(  \beta\right)   &  =\frac{4}{3\pi^{2}}\left(
\frac{{{\beta}}^{\frac{1}{2}}\left(  4{\beta-9}\right)  \Gamma\left(
\frac{\beta-2}{4}\right)  }{4{\Gamma\left(  \frac{\beta}{4}\right)  }}\right.
\nonumber\\
&  \left.  \frac{\alpha_{0}^{2}}{\beta-4}\left(  -1355{{\beta}-128}0{{\beta
}^{2}+}240{{\beta}^{3}}\right)  +\frac{9\pi^{2}\Gamma\left(  \frac{2+\beta}%
{4}\right)  }{{\Gamma\left(  \frac{\beta}{4}\right)  {\beta}}^{\frac{1}{2}}%
}\right)
\end{align}
and Eq.$\left(  \ref{dLB}\right)  $ gives the following results%
\begin{equation}
\left\{
\begin{tabular}
[c]{ccc}\hline
\multicolumn{1}{|c}{$\alpha_{0}$} & \multicolumn{1}{|c}{$\beta_{m}$} &
\multicolumn{1}{|c|}{$\tilde{\Lambda}_{\alpha_{0}}\left(  \beta_{m}\right)  $%
}\\\hline
$1$ & $4.271$ & $988.568$\\
$10$ & $4.384$ & $5952.402$\\
$20$ & $4.395$ & $11443.529$%
\end{tabular}
\ \ \right.  .
\end{equation}
As we can see, the pattern relating the value of $\alpha_{0}$ with the value
of $\tilde{\Lambda}_{\alpha_{0}}\left(  \beta_{m}\right)  $ is valid also in
this case. As concern the flat case $\tilde{\Lambda}_{\alpha_{0}}\left(
\beta\right)  $ reduces to%
\begin{equation}
\tilde{\Lambda}\left(  \beta\right)  =\frac{4}{3\pi^{2}}\left(  \frac{{{\beta
}}^{\frac{1}{2}}\left(  4{\beta-9}\right)  \Gamma\left(  \frac{\beta-2}%
{4}\right)  }{4{\Gamma\left(  \frac{\beta}{4}\right)  }}+\frac{\alpha_{0}^{2}%
}{\beta-4}\left(  -1355{{\beta}-128}0{{\beta}^{2}+}240{{\beta}^{3}}\right)
\right)
\end{equation}
and the minimization procedure gives the following results%
\begin{equation}
\left\{
\begin{tabular}
[c]{ccc}\hline
\multicolumn{1}{|c}{$\alpha_{0}$} & \multicolumn{1}{|c}{$\beta_{m}$} &
\multicolumn{1}{|c|}{$\tilde{\Lambda}_{\alpha_{0}}\left(  \beta_{m}\right)  $%
}\\\hline
$1$ & $4.408$ & $552.721$\\
$10$ & $4.408$ & $5492.789$\\
$20$ & $4.409$ & $10981.754$%
\end{tabular}
\ \ \right.  .
\end{equation}

\section{Conclusions}

\label{p5} In this paper, we have studied the cosmological constant problem
using the deformed WDW equation. Even though we have derived a general
expression for the deformed WDW equation, we have taken into account only the
second and fourth powers of the momentum variable. This deformed WDW equation
was obtained by deforming the second quantized canonical commutation relations
between the field variable and its conjugate momentum. As a consequence, we
have obtained higher order derivative correction terms for the kinetic part of
the WDW equation. It has been demonstrated that these correction terms could
be used to explain the existence of a cosmological constant and we have
observed that the physics of this system depends on the choice of the factor
ordering of the operator. This is usual also in ordinary GR. In this paper, we
have also analyzed the dependence of the cosmological constant on the operator
ordering parameters for various cases. We have analyzed the dependence of the
cosmological constant of the deformation parameter for the generalized
semiclassical case. Note that for generalized semiclassical case we mean the
case in which all the parameters of the factor ordering vanish. As a further
specific case, we have also analyzed flat space case as an example of this
generalized semiclassical case. Finally, we also analyzed another case, where
the factor ordering parameters do not vanish. Unfortunately, we have found a
cosmological constant which is at Planckian scale and not at the present
scale. This could look like a failure of the procedure. Actually this is not
the case. Indeed, the procedure reveals a non vanishing cosmological constant
that should not be there and the most striking fact is that a cosmological
constant is predicted also for flat space, namely the pure GUP distortion is
able to create a VEV. However, how to drive the generated cosmological
constant close to the observed value is a question that can be addressed
including particular potentials like the VSL theory\cite{RemoVSL} or the
Ho\v{r}ava-Lifshitz theory\cite{RemoHL}. However, this goes beyond the scope
of the present paper. It is also interesting to note that the deformation of
the WDW equation also produced a minimum value for the scale factor of the
universe. Thus, the big bang singularity can be avoided. As the deformation of
the WDW equation was related to the existence of the cosmological, it was
argued that the existence of the cosmological constant might prevent the
existence of the big bang singularity. It may be noted that the effect of the
operator ordering on the physics of the system has been studied \cite{01}%
-\cite{10}. In fact, it has been demonstrated that the tunneling wave function
can only be consistently defined for particular choices of operator ordering,
and the no-boundary wave function can be defined independently of operator
ordering \cite{oo}. It would be interesting to repeat this analysis for both
tunneling wave function and no-boundary wave function, by using this deformed
WDW equation.

\appendix{}

\section{Integrals for the Wave Function}

\label{Appe}If the trial wave function assumes the form%
\begin{equation}
\Psi\left(  x\right)  =x^{\frac{\beta-\left(  q+r+s+1\right)  }{2}}\exp\left(
-\frac{\beta x^{4}}{2}\right)  {,}%
\end{equation}
then the first term of the kinetic term becomes%
\begin{gather}
K_{1}=\int_{0}^{\infty}dxx^{q+r+s+p}\Psi^{\ast}\left(  x\right)
\mathcal{H}_{1}\Psi\left(  x\right) \nonumber\\
=-\int_{0}^{\infty}dxx^{q+r+s+p}\Psi^{\ast}\left(  x\right)  \left[
x^{-q}\frac{d}{dx}\left(  x^{q}\frac{d}{dx}\right)  \right]  \Psi\left(
x\right) \nonumber\\
=\frac{{\beta}^{\frac{2-\beta}{4}}}{16}\,\left(  A+4{\beta}\right)
\Gamma\left(  \frac{\beta-2}{4}\right)  , \label{K1}%
\end{gather}
where $\Gamma\left(  x\right)  $ is the gamma function and where we have
defined%
\begin{equation}
A=\left(  {q}-1\right)  ^{2}-8-\left(  p+r+s\right)  ^{2}. \label{A}%
\end{equation}
The second term containing higher order derivatives is%
\begin{align}
K_{2}  &  =\int_{0}^{\infty}dxx^{q+r+s+p}\Psi^{\ast}\left(  x\right)
\mathcal{H}_{2}\Psi\left(  x\right) \nonumber\\
&  =5\alpha_{0}^{2}\int_{0}^{\infty}dxx^{q+r+s+p}\Psi^{\ast}\left(  x\right)
x^{-s}\nonumber\\
&  \times\frac{d}{dx}\left(  x^{s-r}\frac{d}{dx}\left(  x^{r-q}\frac{d}%
{dx}\left(  x^{q}\frac{d}{dx}\right)  \right)  \right)  \Psi\left(  x\right)
\nonumber\\
&  =5\alpha_{0}^{2}\left(  -\frac{B_{1}}{64}{{\beta}^{\frac{4-\beta}{4}}%
-}\frac{B_{2}}{8}{{\beta}^{\frac{8-\beta}{4}}+}\frac{3}{4}{{\beta}%
^{\frac{12-\beta}{4}}}\right)  {\Gamma\left(  \frac{\beta-4}{4}\right)  ,}
\label{K2}%
\end{align}
where we have defined
\begin{align}
B\,_{1}  &  ={p}^{4}+2{p}^{3}q-2{p}^{2}qr-2{p}^{2}qs-2{p}^{2}{r}^{2}-2{p}%
^{2}{s}^{2}\nonumber\\
&  -2p{q}^{3}-4p{q}^{2}r-4p{q}^{2}s-2pq{r}^{2}-4pqrs-2pq{s}^{2}\nonumber\\
&  -{q}^{4}-2{q}^{3}r-2{q}^{3}s-4{q}^{2}rs+2q{r}^{3}-2q{r}^{2}s-2qr{s}%
^{2}\nonumber\\
&  +2q{s}^{3}+{r}^{4}-2{r}^{2}{s}^{2}+{s}^{4}+2{p}^{3}+4{p}^{2}q+2{p}%
^{2}r+6{p}^{2}s\nonumber\\
&  +2p{q}^{2}-2p{r}^{2}-12prs-2p{s}^{2}-2{q}^{2}r-6{q}^{2}s-4q{r}^{2}%
-12q{s}^{2}\nonumber\\
&  -2{r}^{3}+6{r}^{2}s+2r{s}^{2}-6{s}^{3}-8{p}^{2}%
-94\,pq-44\,pr-52\,ps-86\,{q}^{2}\nonumber\\
&  -94\,qr-78\,qs-8\,{r}^{2}-76\,rs+8\,{s}^{2}+14\,p-14\,r-90\,s-105
\label{B1}%
\end{align}
and%
\begin{equation}
B_{2}=\,{3q}\left(  q+{p+r+s}\right)  {\,+2\,pr+2ps+2\,rs-p+r+3\,s+27.}
\label{B2}%
\end{equation}
{The contribution coming from the potential term is composed by%
\begin{equation}
P_{1}=\int_{0}^{\infty}dxx^{1+\beta}\exp\left(  -\beta x^{4}\right)
=\frac{\Gamma\left(  \frac{2+\beta}{4}\right)  }{4{\beta}^{\frac{2+\beta}{4}}}
\label{P1}%
\end{equation}
and%
\begin{equation}
P_{2}=\int_{0}^{\infty}dxx^{3+\beta}\exp\left(  -\beta x^{4}\right)
=\frac{{\Gamma\left(  \frac{\beta}{4}\right)  }}{16{{\beta}^{\frac{\beta}{4}}%
}}{.} \label{P2}%
\end{equation}
}


\begin{thebibliography}{99}                                                                                               %


\bibitem {l}L.~J.~Garay, ``Quantum gravity and minimum length,''
Int.\ J.\ Mod.\ Phys.\ A \textbf{10}, 145 (1995) [gr-qc/9403008].

\bibitem {l2}X.~Calmet, M.~Graesser and S.~D.~H.~Hsu, ``Minimum length from
quantum mechanics and general relativity,'' Phys.\ Rev.\ Lett.\ \textbf{93},
211101 (2004) [hep-th/0405033].

\bibitem {l4}L.~J.~Garay, ``Space-time foam as a quantum thermal bath,''
Phys.\ Rev.\ Lett.\ \textbf{80}, 2508 (1998) [gr-qc/9801024].

\bibitem {l5}L.~J.~Garay, ``Quantum evolution in space-time foam,''
Int.\ J.\ Mod.\ Phys.\ A \textbf{14}, 4079 (1999) [gr-qc/9911002].

\bibitem {z4}D.~Amati, M.~Ciafaloni and G.~Veneziano, ``Can Space-Time Be
Probed Below the String Size?,'' Phys.\ Lett.\ B \textbf{216}, 41 (1989).

\bibitem {1}K.~Konishi, G.~Paffuti and P.~Provero, ``Minimum Physical Length
and the Generalized Uncertainty Principle in String Theory,'' Phys.\ Lett.\ B
\textbf{234}, 276 (1990).

\bibitem {2}T.~Yoneya, ``On the Interpretation of Minimal Length in String
Theories,'' Mod.\ Phys.\ Lett.\ A \textbf{4}, 1587 (1989).

\bibitem {z5}D.~J.~Gross and P.~F.~Mende, ``String Theory Beyond the Planck
Scale,'' Nucl.\ Phys.\ B \textbf{303}, 407 (1988).

\bibitem {loop}T.~Thiemann, ``Lectures on loop quantum gravity,'' Lect.\ Notes
Phys.\ \textbf{631}, 41 (2003)

\bibitem {4}A.~Perez, ``Spin foam models for quantum gravity,''
Class.\ Quant.\ Grav.\ \textbf{20}, R43 (2003)

\bibitem {5}A.~Ashtekar and J.~Lewandowski, ``Background independent quantum
gravity: A Status report,'' Class.\ Quant.\ Grav.\ \textbf{21}, R53 (2004) [gr-qc/0404018].

\bibitem {loop1}P.~Dzierzak, J.~Jezierski, P.~Malkiewicz and W.~Piechocki,
``The minimum length problem of loop quantum cosmology,'' Acta
Phys.\ Polon.\ B \textbf{41}, 717 (2010) [arXiv:0810.3172 [gr-qc]].

\bibitem {nc}M.~R.~Douglas and N.~A.~Nekrasov, ``Noncommutative field
theory,'' Rev.\ Mod.\ Phys.\ \textbf{73}, 977 (2001) [hep-th/0106048].

\bibitem {cn}M.~A.~Gorji, K.~Nozari and B.~Vakili, ``Spacetime singularity
resolution in Snyder noncommutative space,'' Phys.\ Rev.\ D \textbf{89}, no.
8, 084072 (2014) [arXiv:1403.2623 [gr-qc]].

\bibitem {bh}M.~Maggiore, ``A Generalized uncertainty principle in quantum
gravity,'' Phys.\ Lett.\ B \textbf{304}, 65 (1993) [hep-th/9301067].

\bibitem {bh1}M.~I.~Park, ``The Generalized Uncertainty Principle in (A)dS
Space and the Modification of Hawking Temperature from the Minimal Length,''
Phys.\ Lett.\ B \textbf{659}, 698 (2008) [arXiv:0709.2307 [hep-th]].

\bibitem {gu}A.~Kempf, ``Uncertainty relation in quantum mechanics with
quantum group symmetry,'' J.\ Math.\ Phys.\ \textbf{35}, 4483 (1994) [hep-th/9311147].

\bibitem {ug}A.~Kempf, G.~Mangano and R.~B.~Mann, ``Hilbert space
representation of the minimal length uncertainty relation,'' Phys.\ Rev.\ D
\textbf{52}, 1108 (1995) [hep-th/9412167].

\bibitem {1h}H.~Hinrichsen and A.~Kempf, ``Maximal localization in the
presence of minimal uncertainties in positions and momenta,''
J.\ Math.\ Phys.\ \textbf{37}, 2121 (1996) [hep-th/9510144].

\bibitem {h1}A.~Kempf and G.~Mangano, ``Minimal length uncertainty relation
and ultraviolet regularization,'' Phys.\ Rev.\ D \textbf{55}, 7909 (1997) [hep-th/9612084].

\bibitem {1d}S.~Das and E.~C.~Vagenas, ``Universality of Quantum Gravity
Corrections,'' Phys.\ Rev.\ Lett.\ \textbf{101}, 221301 (2008)
[arXiv:0810.5333 [hep-th]].

\bibitem {d1}S.~Das and E.~C.~Vagenas, ``Phenomenological Implications of the
Generalized Uncertainty Principle,'' Can.\ J.\ Phys.\ \textbf{87}, 233 (2009)
[arXiv:0901.1768 [hep-th]].

\bibitem {ds}J.~A.~Magpantay, ``Dual doubly special relativity,''
Phys.\ Rev.\ D \textbf{84}, 024016 (2011) [arXiv:1011.3888 [hep-th]].

\bibitem {sd}H.~Calisto and C.~Leiva, ``Generalized commutation relations and
DSR theories, a close relationship,'' Int.\ J.\ Mod.\ Phys.\ D \textbf{16},
927 (2007) [hep-th/0509227].

\bibitem {li}A.~F.~Ali, S.~Das and E.~C.~Vagenas, ``Discreteness of Space from
the Generalized Uncertainty Principle,'' Phys.\ Lett.\ B \textbf{678}, 497
(2009) [arXiv:0906.5396 [hep-th]].

\bibitem {il}S.~Das, E.~C.~Vagenas and A.~F.~Ali, ``Discreteness of Space from
GUP II: Relativistic Wave Equations,'' Phys.\ Lett.\ B \textbf{690}, 407
(2010) [Phys.\ Lett.\ \textbf{692}, 342 (2010)] [arXiv:1005.3368 [hep-th]].

\bibitem {gt}M.~Kober, ``Gauge Theories under Incorporation of a Generalized
Uncertainty Principle,'' Phys.\ Rev.\ D \textbf{82}, 085017 (2010)
[arXiv:1008.0154 [physics.gen-ph]].

\bibitem {tg}M.~Kober, ``Electroweak Theory with a Minimal Length,''
Int.\ J.\ Mod.\ Phys.\ A \textbf{26}, 4251 (2011) [arXiv:1104.2319 [hep-ph]].

\bibitem {6}M.~Faizal, ``Consequences of Deformation of the Heisenberg
Algebra,'' Int.\ J.\ Geom.\ Meth.\ Mod.\ Phys.\ \textbf{12}, no. 02, 1550022
(2014) [arXiv:1404.5024 [hep-th]].

\bibitem {7}M.~Faizal and S.~I.~Kruglov, ``Deformation of the Dirac
Equation,'' arXiv:1406.2653 [physics.gen-ph].

\bibitem {w}B.~Vakili and H.~R.~Sepangi, ``Generalized uncertainty principle
in Bianchi type I quantum cosmology,'' Phys.\ Lett.\ B \textbf{651}, 79 (2007)
[arXiv:0706.0273 [gr-qc]].

\bibitem {8}M.~Kober, ``Generalized Quantization Principle in Canonical
Quantum Gravity and Application to Quantum Cosmology,''
Int.\ J.\ Mod.\ Phys.\ A \textbf{27}, 1250106 (2012) [arXiv:1109.4629 [gr-qc]].

\bibitem {9}B.~Majumder, ``Dilaton cosmology and the Modified Uncertainty
Principle,'' Phys.\ Rev.\ D \textbf{84}, 064031 (2011) [arXiv:1106.4494 [gr-qc]].

\bibitem {1w}B.~Vakili, ``Cosmology with minimal length uncertainty
relations,'' Int.\ J.\ Mod.\ Phys.\ D \textbf{18}, 1059 (2009)
[arXiv:0811.3481 [gr-qc]].

\bibitem {qb}B.~Majumder, ``Quantum Black Hole and the Modified Uncertainty
Principle,'' Phys.\ Lett.\ B \textbf{701}, 384 (2011) [arXiv:1105.5314 [gr-qc]].

\bibitem {t}M.~Faizal, ``Deformation of Second and Third Quantization,''
Int.\ J.\ Mod.\ Phys.\ A \textbf{30}, no. 09, 1550036 (2015) [arXiv:1503.04797 [gr-qc]].

\bibitem {w1}M.~Faizal, ``Deformation of the Wheeler-DeWitt Equation,''
Int.\ J.\ Mod.\ Phys.\ A \textbf{29}, no. 20, 1450106 (2014) [arXiv:1406.0273 [gr-qc]].

\bibitem {Remo}R.~Garattini, \textquotedblleft The Cosmological constant as an
eigenvalue of a Sturm-Liouville problem and its
renormalization,\textquotedblright\ J.\ Phys.\ A \textbf{39}, 6393 (2006)
[gr-qc/0510061]. R.~Garattini, \textquotedblleft Extracting the cosmological
constant from the Wheeler-DeWitt equation in a modified gravity
theory,\textquotedblright\ J.\ Phys.\ A \textbf{41}, 164057 (2008)
[arXiv:0712.3246 [gr-qc]].

\bibitem {RGGM}R.~Garattini and G.~Mandanici, \textquotedblleft Modified
Dispersion Relations lead to a finite Zero Point Gravitational
Energy,\textquotedblright\ \textsl{Phys.\ Rev.}\ \textbf{D 83}, 084021 (2011)
[arXiv:1102.3803 [gr-qc]].

\bibitem {RGf(R)}R. Garattini, \textquotedblleft Distorting General
Relativity: Gravity's Rainbow and f(R) theories at
work\textit{{\textquotedblright,}} \textsl{JCAP} 1306 (2013)\textbf{ 017};
[arXiv:1210.7760 [gr-qc]]. S.~Capozziello and R.~Garattini, \textquotedblleft
The cosmological constant as an eigenvalue of f(R)-gravity Hamiltonian

constraint\textquotedblright,Class.\ Quant.\ Grav.\ \textbf{24}, 1627 (2007) [arXiv:gr-qc/0702075].

\bibitem {Vilenkin}A.\ Vilenkin, \textsl{Phys.\ Rev. }\textbf{D 37} (1988) 888.

\bibitem {e}R.~Garattini, \textquotedblleft Extracting the Maxwell charge from
the Wheeler-DeWitt equation\textquotedblright,\ \textsl{Phys.\ Lett.}%
\ \textbf{B} \textbf{666}, 189 (2008) [arXiv:0807.0082 [gr-qc]].

\bibitem {m}R.~Garattini and B.~Majumder, \textquotedblleft Electric Charges
and Magnetic Monopoles in Gravity's Rainbow\textquotedblright%
,\ \textsl{Nucl.\ Phys.}\ \textbf{B} \textbf{883}, 598 (2014) [arXiv:1305.3390 [gr-qc]].

\bibitem {Naked}R.~Garattini and B.~Majumder, \textquotedblleft Naked
Singularities are not Singular in Distorted Gravity\textquotedblright%
,\ \textsl{Nucl.\ Phys.}\ \textbf{B} \textbf{884}, 125 (2014) [arXiv:1311.1747 [gr-qc]].

\bibitem {RemoHL}R. Garattini, \textquotedblleft The Cosmological constant as
an eigenvalue of the Hamiltonian constraint in Ho\v{r}ava-Lifshits
theory\textquotedblright, \textsl{Phys. Rev.} \textbf{D 86} 123507 (2012);
[arXiv:0912.0136 [gr-qc]].

\bibitem {OBCZ}O. Bertolami and C. A. D. Zarro, \textquotedblleft
Ho\v{r}ava-Lifshitz Quantum Cosmology\textquotedblright, \textsl{Phys. Rev.}
\textbf{D 84} 044042 (2011); [arXiv:1106.0126 [hep-th]].

\bibitem {RemoVSL}R. Garattini and M. De Laurentis, \textquotedblleft The
Cosmological Constant as an Eigenvalue of the Hamiltonian constraint in a
Varying Speed of Light theory\textquotedblright\textit{, }arXiv:1503.03677 [gr-qc].

\bibitem {cosm}A.~G.~Riess \textit{et al.} [Supernova Search Team
Collaboration], ``Observational evidence from supernovae for an accelerating
universe and a cosmological constant,'' Astron.\ J.\ \textbf{116}, 1009 (1998) [astro-ph/9805201].

\bibitem {n2}A.~G.~Riess, A.~V.~Filippenko, W.~Li and B.~P.~Schmidt, ``An
indication of evolution of type ia supernovae from their risetimes,''
Astron.\ J.\ \textbf{118}, 2668 (1999) [astro-ph/9907038].

\bibitem {n1}A.~G.~Riess \textit{et al.} [Supernova Search Team
Collaboration], ``The farthest known supernova: support for an accelerating
universe and a glimpse of the epoch of deceleration,''
Astrophys.\ J.\ \textbf{560}, 49 (2001) [astro-ph/0104455].

\bibitem {cosm1}S.~Perlmutter \textit{et al.} [Supernova Cosmology Project
Collaboration], ``Discovery of a supernova explosion at half the age of the
Universe and its cosmological implications,'' Nature \textbf{391}, 51 (1998) [astro-ph/9712212].

\bibitem {01}D.~L.~Wiltshire, ``Wave functions for arbitrary operator ordering
in the de Sitter minisuperspace approximation,''
Gen.\ Rel.\ Grav.\ \textbf{32}, 515 (2000) [gr-qc/9905090].

\bibitem {10}R.~Steigl and F.~Hinterleitner, ``Factor ordering in standard
quantum cosmology,'' Class.\ Quant.\ Grav.\ \textbf{23}, 3879 (2006) [gr-qc/0511149].

\bibitem {oo}N.~Kontoleon and D.~L.~Wiltshire, ``Operator ordering and
consistency of the wave function of the universe,'' Phys.\ Rev.\ D
\textbf{59}, 063513 (1999) [gr-qc/9807075].
\end{thebibliography}
\end{document}